\documentclass[twocolumn,showpacs,preprintnumbers,prl]{revtex4}

\usepackage{citesort}
\usepackage{epsfig}

\begin{document}

\def \d {{\rm d}}

\preprint{BNL-NT-03/32}
\preprint{RBRC-359}

\title{QCD hard-scattering and the sign of the spin 
asymmetry $\mathbf{A_{\mathrm{LL}}^{\pi}}$}
\author{Barbara J\"{a}ger}
\affiliation{Institut f{\"u}r Theoretische Physik, 
Universit{\"a}t Regensburg, D-93040 Regensburg, Germany}
\author{Stefan Kretzer}
\affiliation{Physics Department, Brookhaven National Laboratory,
Upton, New York 11973, USA, and\\ 
RIKEN-BNL Research Center, Brookhaven National Laboratory,
Upton, New York 11973, USA}
\author{Marco Stratmann}
\affiliation{Institut f{\"u}r Theoretische Physik, 
Universit{\"a}t Regensburg, D-93040 Regensburg, Germany}
\author{Werner Vogelsang}
\affiliation{Physics Department, Brookhaven National Laboratory,
Upton, New York 11973, USA, and\\ 
RIKEN-BNL Research Center, Brookhaven National Laboratory,
Upton, New York 11973, USA}

\begin{abstract}
Recent preliminary {\sc{Phenix}} data are consistent with a negative and 
sizable longitudinal double-spin asymmetry $A_{\mathrm{LL}}^{\pi}$
for $\pi^0$ production at moderate transverse momentum 
$p_\perp \simeq 1 \div 4\, {\rm GeV}$ and central rapidity. By means of 
a systematic investigation of the relevant degrees of freedom we show 
that the perturbative QCD framework at leading power in $p_{\perp}$ 
produces at best a very small negative asymmetry in this kinematic range. 
\end{abstract}

\pacs{13.88+e, 12.38.Bx, 13.85.Ni}

\maketitle

\section{Introduction}
The determination of the nucleon's polarized gluon density
is a major goal of current experiments with longitudinally
polarized protons at RHIC~\cite{rhicrev}. 
It can be accessed through measurement
of the spin asymmetries
\begin{equation}
\label{eq:asydef}
A_{\mathrm{LL}}=\frac{\d\Delta \sigma}{\d\sigma}=\frac{ 
\d\sigma^{++} - \d\sigma^{+-}}{\d\sigma^{++} + \d\sigma^{+-}}
\end{equation}
for high transverse momentum ($p_{\perp}$) reactions such as 
$pp\to \pi X$, jet$X$, $\gamma X$. 
In Eq.~(\ref{eq:asydef}), $\sigma^{++}$ ($\sigma^{+-}$) denotes
the cross section for scattering of two protons with same (opposite)
helicities. High $p_{\perp}$ implies large momentum transfer, and the cross
sections for such reactions may be factorized into
long-distance pieces that contain the desired information on the
(spin) structure of the nucleon, and short-distance parts that
describe the hard interactions of the partons and are amenable to
QCD perturbation theory (pQCD). All of the reactions listed above have
partonic Born cross sections involving gluons in the initial
state and may therefore serve to examine the gluon content
of the scattering longitudinally polarized protons.

In this paper, we will consider the spin asymmetry 
$A_{\mathrm{LL}}^{\pi}$ for high-$p_{\perp}$ $\pi^0$ production, 
for which very recently the {\sc{Phenix}} collaboration has presented 
first preliminary data~\cite{phenix} at center-of-mass (c.m.s.)
energy $\sqrt{S}=200$~GeV and central rapidity. 
The data are consistent with a sizable (up to a few per cent)
negative $A_{\mathrm{LL}}^{\pi}$ in the region $p_\perp \sim 1 \div 4$~GeV. 
Even though the experimental uncertainties are still large
and leave room for a different behavior of $A_{\mathrm{LL}}^{\pi}$,
the new data motivate us to entertain the unexpected
possibility of $A_{\mathrm{LL}}^{\pi}$ being negative.
To our knowledge, there have been no predictions of a
substantially negative $A_{\mathrm{LL}}^{\pi}$ in the literature.
This is in itself interesting, also because at moderate $p_\perp$
the spin asymmetry $A_{\mathrm{LL}}^{\pi}$ is expected to be particularly
sensitive to gluon polarization.

\section{Hard-Scattering calculation}
We may write the polarized high-$p_\perp$ cross section as 
\begin{eqnarray} \label{eq:crosec}
\frac{\d\Delta \sigma^{\pi}}{\d p_\perp \d \eta} &=&\sum_{a,b,c}\, 
\int dx_a \int dx_b \int dz_c \,\,
\Delta a (x_a,\mu) \,\Delta b (x_b,\mu) \nonumber \\ [2mm]
&\times& \frac{\d\Delta \hat{\sigma}_{ab}^{c}
(p_\perp, \eta, x_a, x_b, z_c, \mu)}
{\d p_\perp \d \eta}\, D_c^{\pi}(z_c,\mu) \,  \; ,
\end{eqnarray}
where $\eta$ is the pion's pseudorapidity. The $\Delta a,\Delta b \;(
a,b=q,\bar{q},g)$ are the polarized parton densities; for instance, 
\begin{equation}
\label{eq:pdf}
\Delta g(x,\mu) \equiv g_+(x,\mu) -
                       g_-(x,\mu) \; ,
\end{equation}
(the sign referring to the gluon helicity in a proton of positive
helicity) is the polarized gluon distribution. 
The sum in Eq.~(\ref{eq:crosec}) is over all partonic channels $a+b\to
c + X$, with their associated polarized cross sections
$d\Delta \hat{\sigma}_{ab}^{c}$. These start at ${\cal{O}}(\alpha_s^2)$ 
in the strong coupling with the QCD tree-level scatterings:
(i)~$g g \rightarrow gg$, (ii)~$g g \rightarrow q {\bar q}$, 
(iii)~$gq (\bar{q}) \rightarrow g q (\bar{q})$,
(iv)~$q {\bar q} \rightarrow q {\bar q} $, $q {\bar q} 
\rightarrow g g $, $qq\to qq$, $qq'\to qq'$,
$q\bar{q}\rightarrow q'\bar{q}'$. 
The transition of parton $c$ into the observed $\pi^0$
is described by the (spin-independent) fragmentation function
$D_c^{\pi}$. 
The functions in Eq.~(\ref{eq:crosec}) are
tied together by their dependence on the factorization/renormalization
scale $\mu$ which is of the order of the hard scale $p_\perp$,
but not further specified. 
All next-to-leading order [NLO, ${\cal O}(\alpha_s^3)$] QCD
contributions to polarized parton scattering are known \cite{JSSV}. 
Corrections to Eq.~(\ref{eq:crosec}) itself are
down by inverse powers of $p_\perp$ and are thus expected 
to become relevant if $p_\perp$ is not much bigger than typical
hadronic mass scales. We neglect such contributions for now, but
will briefly return to them later. 

\begin{figure}[t]
\begin{center}
\vspace*{-0.6cm}
\epsfig{figure=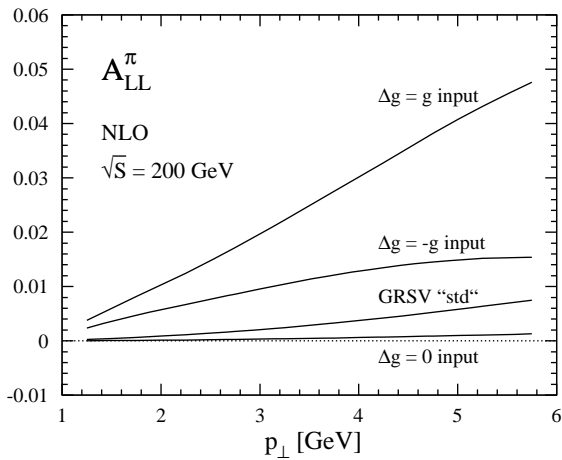,width=0.48\textwidth}
\end{center}
\vspace*{-0.7cm}
\caption{NLO predictions for $A_{\mathrm{LL}}^{\pi}$
based on different assumptions about $\Delta g$ at the input scale
for the evolution in \cite{grsv}. 
\label{fig:all}}
\vspace*{-0.5cm}
\end{figure}

To set the stage for our further considerations, Fig.~\ref{fig:all} 
shows NLO predictions for $A_{\mathrm{LL}}^{\pi}$,
for various gluon polarizations $\Delta g$, all proposed within the
framework of the analysis of data from polarized deeply-inelastic 
scattering (DIS) in~\cite{grsv}. Despite the fact that the
$\Delta g$'s used in Fig.~\ref{fig:all} are all very different from 
one another, none of the resulting $A_{\mathrm{LL}}^{\pi}$ is
negative in the $p_{\perp}$ region we display. 
%
\section{Basic observations} 
As a first step in our more detailed analysis of $A_{\mathrm{LL}}^{\pi}$
we discuss the various ingredients to Eq.~(\ref{eq:crosec}), 
other than the unknown $\Delta g$ that one hopes to probe: the 
$\Delta q$ and $\Delta \bar{q}$ distributions, the calculated partonic
cross sections, and the fragmentation functions $D_c^{\pi}$. 
They all jointly determine the analyzing
power for $\Delta g$ provided by $A_{\mathrm{LL}}^{\pi}$.

To a first approximation, the $\Delta q$, $\Delta \bar{q}$ 
may be treated as ``known''. $\Delta q + \Delta \bar{q}$ has been 
probed extensively in polarized DIS. There certainly remains 
much room for improvement in our knowledge about them, but this
does not affect $A_{\mathrm{LL}}^{\pi}$ too strongly. To give one
example: for $\pi^0$ production the $qg$ channel depends on
the sum $\Delta u+\Delta \bar{u}+\Delta d+\Delta \bar{d}$, so 
that uncertainties relating to the SU(2) flavor structure of the polarized 
sea are not relevant. 

We now turn to the partonic cross sections and fragmentation
functions. We focus on the lowest order (LO) 
which is expected to capture the qualitatively important points. 
Among the reactions (i)-(iii) listed above that have gluons in the initial 
state, process (ii) has a negative {\em partonic} spin asymmetry
${\hat a}_{\mathrm{LL}}\equiv -1$, while (i) and (iii) both have 
${\hat a}_{\mathrm{LL}}>0$~\cite{rhicrev}. Also, for the $p_{\perp}$ 
region of interest, one has an average $\langle z_c \rangle \sim 0.3\div 0.4$, 
where quarks are more favored than gluons to fragment into pions.
A first guess is, then, to attribute a negative
$A_{\mathrm{LL}}^{\pi}$ to the negative $gg\to q\bar{q}$ cross
section. However, this expectation is refuted by the numerical 
hierarchy in the partonic cross sections: at $\hat{\eta}=0$ in the 
partonic c.m.s., which is most relevant for 
the {\sc{Phenix}} data, channel (i) is (in absolute magnitude) 
larger than (ii) by a factor of about 160. Therefore, if one 
wanted to suppress the (positive) $gg\to gg$ contribution, one 
would effectively have to switch off the gluon-to-pion fragmentation 
function $D_g^{\pi}$. Even though our knowledge of $D_g^{\pi}$ is incomplete, this 
does not appear to be a sensible solution, for two reasons. First, 
data from $e^+e^-$ collisions, in particular analyses of hadron 
production in $b\bar{b}$ plus (mostly gluon) jet final states 
\cite{lepbbg}, do constrain $D_g^{\pi}$ significantly, even at
fairly large $z_c$. For example, the $D_g^{\pi}$'s in the sets of~\cite{kkp} 
and~\cite{kretzer} are in reasonable agreement with these
data, with the one of~\cite{kretzer} arguably setting a lower
bound on $D_g^{\pi}$. Second, elimination of the $gg\to gg$ 
channel would also strongly affect the unpolarized cross section
and reduce it by up to an order of magnitude at RHIC energies.
However, previous {\sc{Phenix}} measurements \cite{phenixunpol} of the 
unpolarized cross section for $pp\to \pi^0 X$ at $\sqrt{S}=200\,
\mathrm{GeV}$ were found to be in excellent agreement with NLO 
calculations employing the $D_c^{\pi}$ of~\cite{kkp}.

We therefore exclude that the $gg\to q\bar{q}$ channel
is instrumental in making $A_{\mathrm{LL}}^{\pi}$ negative, and 
we thus have to investigate possibilities within $\Delta g$ itself, 
and its involvement in $gg\to gg$ and $qg\to qg$ scattering. 
Given that the polarized scattering cross sections
for these reactions are both positive, and that the first process
comes roughly with the square of $\Delta g$, it is immediately
clear that a sizable negative asymmetry will not be easily 
obtained. In the next section, we will demonstrate this for the
particularly instructive case of rapidity integrated cross
sections.

\section{A lower bound on $\mathbf{A_{\mathrm{LL}}^{\pi}}$}
We consider the LO cross section integrated
over all rapidities $\eta$. 
This is not immediately 
relevant for comparison to experiment,
but it does capture the main point 
we want to make. 
From Eq.~(\ref{eq:crosec}), one obtains: 
\begin{eqnarray}  \label{crosec1}
\frac{p_{\perp}^3 \d\Delta \sigma^{\pi}}{\d p_{\perp}} &=& \sum_{a,b,c} 
\int_{x_T^2}^1 \, dx_a \,\Delta a (x_a,\mu)\, 
\int_{\frac{x_T^2}{x_a}}^1  
\, dx_b \,\Delta b (x_b,\mu) \nonumber \\ 
&\times& \int_{\frac{x_T}{\sqrt{x_a x_b}}}^1  \, dz_c\,
D_c^{\pi} (z_c,\mu)\,
\frac{p_{\perp}^3 \d\Delta\hat{\sigma}_{ab}^c}{\d p_{\perp}}
(\hat{x}_T^2,\mu) \; ,
\end{eqnarray}
where $\hat{x}_T^2 \equiv x_T^2/z_c^2 x_a x_b$, $x_T\equiv 2p_{\perp}/\sqrt{S}$. 
It is then convenient to take Mel\-lin moments in $x_T^2$ of the cross section, 
\begin{equation} \label{doublemom}
\Delta\sigma^{\pi} (N) \equiv
\int_0^1 dx_T^2 \left( x_T^2 \right)^{N-1} 
\frac{p_{\perp}^3 \d\Delta\sigma^{\pi}}{\d p_{\perp}} \; .
\end{equation}
One obtains  (we suppress the scale $\mu$ from now on):
\begin{equation} \label{crosec2}
\Delta\sigma^{\pi} (N) = \sum_{a,b,c} \,\Delta a^{N+1}\,
\Delta b^{N+1}\,\Delta
\hat{\sigma}_{ab}^{c,N}\,D_c^{\pi,2N+3}\; ,
\end{equation}
where the $\Delta\hat{\sigma}_{ab}^{c,N}$ are the $\hat{x}_T^2$-moments 
of the partonic cross sections and, as usual,
$ f^N\equiv \int_0^1 dx\, x^{N-1} f(x)$
for the parton distribution and fragmentation functions. 
We now rewrite Eq.~(\ref{crosec2}) in a 
form that makes the dependence on the moments $\Delta g^N$ explicit:
\begin{equation} \label{quad1}
\Delta\sigma^{\pi} (N) =
\left(\Delta g^{N+1} \right)^2 {\cal A}^N + 
2 \Delta g^{N+1} {\cal B}^N + {\cal C}^N \; .
\end{equation}
Here, ${\cal A}^N$ represents the contributions from $gg\to gg$ and
$gg\to q\bar{q}$, ${\cal B}^N$ the ones from $qg\to qg$, and
${\cal C}^N$ those from the (anti)quark scatterings (iv) above; in each 
case, the appropriate combinations of $\Delta q$, $\Delta \bar{q}$ 
distributions and fragmentation functions are included.  

Being a quadratic form in $\Delta g^{N+1}$, $\Delta\sigma^{\pi} (N)$
possesses an extremum, given by the condition 
\begin{equation} \label{dgmin}
{\cal A}^N \Delta g^{N+1}  = -{\cal B}^N \; .
\end{equation}
We note in passing that the same equation may also be derived directly 
from Eq.~(\ref{crosec1}) by regarding the cross section as a
functional of $\Delta g$, using a variational approach, and
taking Mellin moments of the resulting stationarity condition.
In the following we neglect the contribution from the 
$gg\to q\bar{q}$ channel which, as we discussed 
above, is much smaller than that from $gg\to gg$
for the $p_{\perp}$ we are interested in. The
coefficient ${\cal A}^N$ is then positive, and
Eq. (\ref{dgmin}) describes a minimum of $\Delta\sigma^{\pi} (N)$,
with value
\begin{equation} \label{crsecmin}
\Delta\sigma^{\pi} (N) \Big|_{\mathrm{min}} =
 -\left({\cal B}^N \right)^2/{\cal A}^N + {\cal C}^N \; .
\end{equation} 
It is then straightforward to perform a numerical Mellin
inversion of this minimal cross section:
\begin{equation} \label{inverse}
\frac{p_{\perp}^3 \d\Delta {\sigma}^{\pi}}{\d p_{\perp}}
\Bigg|_{\mathrm{min}} = \frac{1}{2\pi i}
\int_{\Gamma} dN \left(x_T^2\right)^{-N} \, \Delta\sigma^{\pi} (N) 
\Big|_{\mathrm{min}}\,, 
\end{equation}
where $\Gamma$ denotes a suitable contour in complex-$N$ space.

\begin{figure}[t]
\begin{center}
\vspace*{-0.6cm}
\epsfig{figure=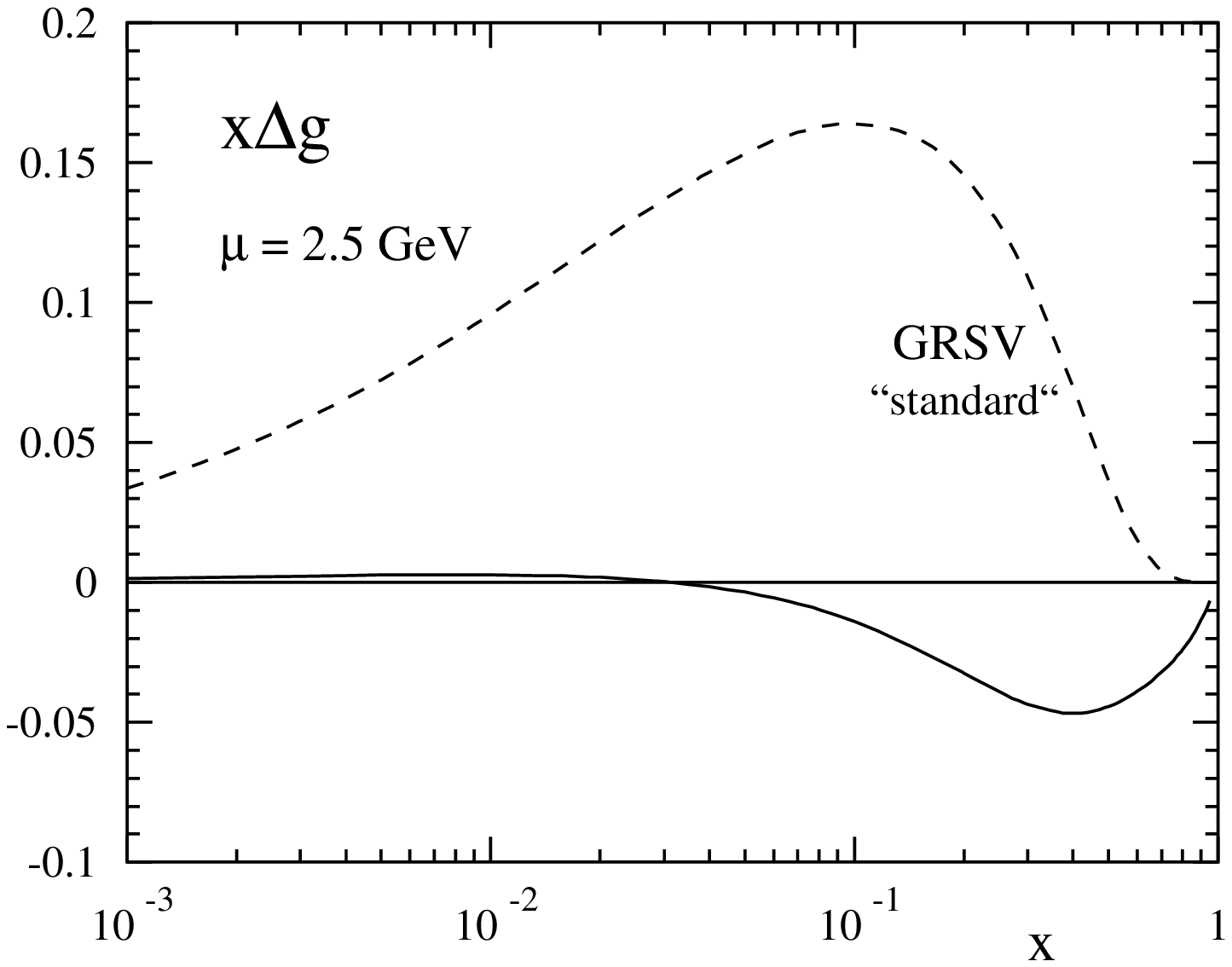,width=0.50\textwidth}
\end{center}
\vspace*{-0.8cm}
\caption{$\Delta g(x,\mu=2.5\,\mathrm{GeV})$ resulting 
from Eq.~(\ref{dgmin}) (solid). The dashed line shows the 
GRSV LO ``standard'' $\Delta g$ \cite{grsv}. \label{fig:dg}}
\vspace*{-0.7cm}
\begin{center}
\epsfig{figure=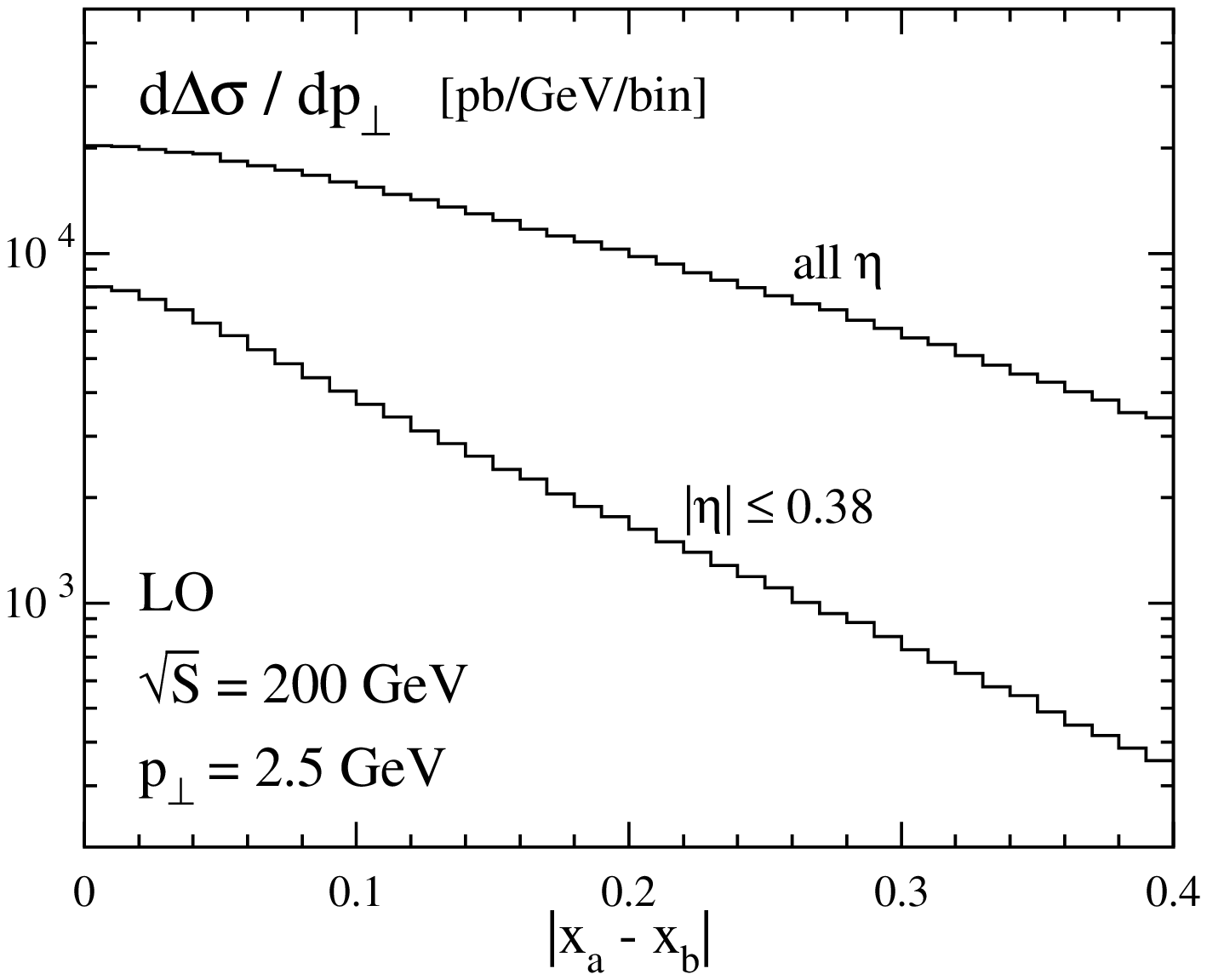,width=0.50\textwidth}
\end{center}
\vspace*{-0.8cm}
\caption{$\d\Delta \sigma^{\pi}/\d p_{\perp}$ in bins of 
$|x_a-x_b|$. \label{fig:x1x2}}
\vspace*{-0.5cm}
\end{figure}

For the numerical evaluation we use the LO $\Delta q$, 
$\Delta \bar{q}$ of GRSV~\cite{grsv}, the $D_c^{\pi}$ of~\cite{kkp},
and a fixed scale $\mu=2.5$~GeV. We find that the minimal 
asymmetry resulting from this exercise is negative indeed, 
but very small: in the range $p_\perp \sim 1 \div 4$~GeV
its absolute value does not exceed $10^{-3}$. The $\Delta g$
in Eq.~(\ref{dgmin}) that minimizes the asymmetry
is shown in Fig.~\ref{fig:dg}, compared to $\Delta g$
of the GRSV LO ``standard'' set \cite{grsv}. One can see that 
it has a node and is generally much smaller than the GRSV one, except 
at large $x$. The node makes it possible to probe the two gluon densities 
in the $gg$ term at values of $x_a$, 
$x_b$ where they have different sign, which helps in decreasing 
$A_{\mathrm{LL}}^{\pi}$. 

Even though we have made some approximations in deriving the bound
in Eq.~(\ref{inverse}), we
do believe that it exhibits the basic difficulty with a 
sizable negative $A_{\mathrm{LL}}^{\pi}$ at moderate
$p_\perp$: the fact that the cross section
is a quadratic form in $\Delta g$ effectively means that
it is bounded from below. Note that this bound does
not {\em always} imply that a negative $A_{\mathrm{LL}}^{\pi}$ is small:
at higher $p_{\perp}$ it does allow a fairly large $A_{\mathrm{LL}}^{\pi}$.
 
In its details, the qualitative picture drawn by our study is of course 
subject to a number of corrections. First of all, we have
integrated the cross section over all rapidities, whereas for the 
{\sc{Phenix}} data $|\eta| \leq 0.38$. It is instructive to investigate the 
qualitative differences associated with this. Figure~\ref{fig:x1x2}
shows the polarized LO cross section at $p_{\perp}=2.5$~GeV versus the 
``distance'' $|x_a-x_b|$ in parton momentum fractions 
in Eq.~(\ref{eq:crosec}), for integration over all $\eta$ and for
$|\eta|\leq 0.38$. Here we have used the GRSV polarized parton densities. 
The larger the rapidity range probed, the more likely become collisions of 
partons with rather different momentum fractions. Indeed, the distribution 
for $|\eta|\leq 0.38$ is steeper, implying that a node in 
$\Delta g(x,\mu)$ will now be somewhat less efficient in promoting 
negative values for the asymmetry.

In a more realistic calculation one would also prefer $\mu\sim p_{\perp}$ 
to a fixed $\mu$. Furthermore, since the $\Delta q$ and $\Delta \bar{q}$ 
are coupled to $\Delta g$ via evolution, any change 
of $\Delta g$ will require a re-tuning of the 
$\Delta q$, $\Delta \bar{q}$ densities, so that the agreement with the
polarized DIS data remains intact. Inclusion of the NLO 
corrections is important as well. 

All these points can be thoroughly addressed 
only in a ``global'' NLO analysis of the data, taking
into account the results from polarized DIS as well. We will 
now report on such an analysis. Given that the data are still
preliminary, this may seem premature. Our primary 
goal, however, is to investigate 
whether the findings of our somewhat idealized case, as summarized by
Eqs.~(\ref{crsecmin}) and (\ref{inverse}),
hold true in general. 

\section{``Global'' NLO analysis}
The main technical difficulty in a full global NLO analysis
of polarized DIS and $pp\to\pi^0 X$ data is the 
numerical complexity of the evaluation of the NLO corrections 
for the latter cross section. A convenient way to alleviate
this problem was presented in \cite{svmell}. Starting from
Eq.~(\ref{eq:crosec}), one expresses the $\Delta a$, $\Delta b$ by 
their Mellin inverses, e.g., 
\begin{equation} \label{eq6}
 \Delta a(x,\mu) = \frac{1}{2 \pi i} \int_{\Gamma_N} dN \; 
x^{-N} \Delta a^N(\mu) \; .
\end{equation}
After interchange of integrations one obtains
\begin{eqnarray} \label{eq:crosecmell}
\frac{\d\Delta \sigma^{\pi}}{\d p_\perp \d \eta} &=&\sum_{a,b}\, 
\int_{\Gamma_N} dN \; 
\int_{\Gamma_M} dM\;\Delta a^N(\mu)\,\Delta b^M(\mu)
\nonumber \\ 
&\times& \rho_{ab}^{\pi} (N,M,p_\perp, \eta, \mu)\; ,
\label{eq8}
\end{eqnarray}
where the $\rho_{ab}^{\pi}$ contain the partonic cross sections,
the fragmentation functions, and all integrations over momentum
fractions, with the factors $x_a^{-N}$ and $x_b^{-M}$ as complex 
``dummy'' parton distributions according to Eq.~(\ref{eq6}). 
The strength of this approach is that there is no dependence 
of $\rho_{ab}^{\pi}$ on the moments $\Delta a^N$, $\Delta b^M$
of the true parton densities. This means that the $\rho_{ab}^{\pi}$ 
can be pre-calculated {\em prior} to the analysis on a specific array
of the two Mellin variables $N$ and $M$. One chooses a convenient
functional form for the parton distributions, depending on a set of 
free parameters. The latter are then determined from a $\chi^2$ 
minimization procedure. The double inverse Mellin transformation 
which finally links the parton distributions with the pre-calculated 
$\rho_{ab}^{\pi}$ of course still needs to be performed in each step 
of the fitting procedure, but becomes extremely fast by  choosing the 
values for $N,M$ on the contours $\Gamma_N$, $\Gamma_M$ simply as the 
supports for a Gaussian integration. 

Following these lines, we have performed a simultaneous analysis
of all data from polarized DIS and of the preliminary {\sc{Phenix}}
data for $pp\to\pi^0 X$. We have used several different
functional forms for the polarized gluon density, in particular 
allowing it to have a node. The quark densities were allowed
to vary as well. We have artificially decreased 
the error bars of the data points for $A_{\mathrm{LL}}^{\pi}$
in order to see whether the fit can be forced to reproduce
a negative $A_{\mathrm{LL}}^{\pi}$ of about $-2\%$ in the
region $p_\perp \simeq 1 \div 4\, {\rm GeV}$. We have also slightly 
shifted individual data points to study the response of the
fit. In no case have we been able to find a fit that yielded
a negative $A_{\mathrm{LL}}^{\pi}$ with absolute value 
larger than a few times $10^{-3}$. Even those fits, however, gave
a negative $A_{\mathrm{LL}}^{\pi}$ only at the higher end
of the $p_\perp$ interval, and invariably they led to a 
polarized gluon density that had a node and tended to violate
positivity $|\Delta g|\leq g$ in certain ranges of $x$. 
The ``global'' analysis thus confirms our qualitative
finding above that any negative $A_{\mathrm{LL}}^{\pi}$ is 
also very small.\\

\section{Conclusions}
Our analysis demonstrates that pQCD at {\em leading power} 
in $p_{\perp}$ predicts that $A_{\mathrm{LL}}^{\pi}$ is bounded from 
below by $A_{\mathrm{LL}}^{\pi}\gtrsim {\cal{O}}(-10^{-3})$
in the region $p_\perp \simeq 1 \div 4 \;{\rm GeV}$. 
The observation relies on collinear factorization 
and on exploring the physically acceptable ranges of
parton distribution and fragmentation functions.

For now, the data \cite{phenix} do not allow a compelling
conclusion on whether the bound is violated or not.
What should one conclude if future, more precise, data
will indeed confirm a sizable negative $A_{\mathrm{LL}}^{\pi}$?
As indicated earlier, corrections to Eq.~(\ref{eq:crosec})
are associated with power-suppressed contributions to the 
cross section. 
Since $p_\perp$ is not
too large, such contributions might well be significant. 
On the other hand, comparisons of unpolarized $\pi^0$ spectra 
measured at colliders with NLO QCD calculations do not exhibit 
any compelling trace of non-leading power effects even down to fairly 
low $p_\perp \gtrsim 1\ {\rm GeV}$, within the uncertainties of the
calculation. It is conceivable that the spin-dependent cross section 
with its fairly tedious cancelations has larger power-suppressed 
contributions than the unpolarized one. One may attempt to model the
effects by implementing an ``intrinsic'' transverse-momentum ($k_{\perp}$)
smearing for the initial partons which generically leads to corrections by
powers of $\langle k_{\perp} \rangle/p_{\perp}$, with $\langle k_{\perp} 
\rangle$ an average $k_{\perp}$. Such effects were shown to have
indeed some potential impact on $A_{\mathrm{LL}}^{\pi}$ at 
$p_{\perp}\leq 5$~GeV~\cite{vw}. A negative 
$A_{\mathrm{LL}}^{\pi}$
would open up a quite unexpected window on 
aspects of nucleon structure and
limitations of pQCD so far little explored.

\acknowledgments
We thank A.\ Bazilevsky, 
G.\ Bunce, A.\ Deshpande, E.\ Leader, N.\ Saito, A.\ Sch\"{a}fer, 
and J.\ Soffer for useful discussions. 
This work was supported in part by BMBF and DFG, Berlin/Bonn
and by  RIKEN, BNL, 
and the U.S.\ Department of Energy (contract DE-AC02-98CH10886).


\end{document}